\documentclass[%
 reprint,
superscriptaddress,
 amsmath,amssymb,
 aps,
]{revtex4-2}

\usepackage{graphicx}
\usepackage[utf8]{inputenc}
\usepackage{dcolumn}
\usepackage{bm}
\usepackage{float}
\usepackage{hyperref}
\hypersetup{
    colorlinks=true,
    linkcolor=blue,
    filecolor=magenta,      
    urlcolor=blue,
    pdftitle={Overleaf Example},
    pdfpagemode=FullScreen,
    }

\begin{document}

\preprint{APS/123-QED}

\title{Wavelength scaling and multi-color operation  of a plasma driven attosecond X-ray source via harmonic generation}

\author{Rafi Hessami}
 \affiliation{SLAC National Accelerator Laboratory}
 \email{rafimah@stanford.edu}

\author{Jenny Morgan}
\affiliation{SLAC National Accelerator Laboratory}
\email{jmorgan@slac.stanford.edu}

\author{River Robles}
\affiliation{SLAC National Accelerator Laboratory}
\email{riverr@stanford.edu}

\author{Kirk A. Larsen}
\affiliation{SLAC National Accelerator Laboratory}
\affiliation{Stanford PULSE Institute}
\email{larsenk@stanford.edu}

 \author{Agostino Marinelli}
 \affiliation{SLAC National Accelerator Laboratory}
 \email{marinelli@stanford.edu}


\author{Claudio Emma}
 \affiliation{SLAC National Accelerator Laboratory}%
 \email{cemma@slac.stanford.edu}


\date{\today}

\begin{abstract}

The generation of high power coherent soft X-ray pulses of sub-100 as duration and 10 nm wavelength using beams from a GeV energy plasma wakefield accelerator has been recently investigated in Ref. \cite{Emma2021}. As a future upgrade to this concept, this contribution investigates scaling to shorter X-ray wavelengths by cascading undulators tuned to higher harmonics of the fundamental. We present two simulation studies for plasma-driven attosecond harmonic generation schemes with final photon wavelengths of 2 nm and 0.40 nm. We demonstrate in these schemes that using undulators with retuned fundamental frequencies can produce GW-scale pulses of sub-nm radiation with tens of attosecond-scale pulse lengths, an order of magnitude shorter than current state-of-the-art attosecond XFELs. This multi-pulse multi-color operation will be broadly applicable to attosecond pump-probe experiments.


\end{abstract}

\maketitle


\section{Introduction}







X-ray Free Electron Lasers (XFELs) are powerful scientific instruments that enable the exploration of natural phenomena down to attosecond timescales and Angstrom spatial scales \cite{Pellegrini2016}. 
State-of-the-art attosecond XFELs are able to produce X-ray beams with 100 $\mu$J-scale pulse energies with measured pulse lengths down to 280 as at nm-wavelengths \cite{Duris, Duris2021, Zhang2020}. Single-spike hard X-ray operation has also been demonstrated with estimated pulse durations in the hundreds of attoseconds \cite{huang2017generating, marinelli2017experimental, PhysRevResearch.2.042018}.
Reaching sub-100-as timescale has interesting ramifications for studying ultrafast electronic phenomena, such as allowing the study of molecular relaxation and charge redistribution following the sudden removal of an electron, or improving the fidelity of attosecond pump/probe experiments for the observation of charge migration in molecules \cite{Krausz2009, Breidbach2005}. To reach the sub-100 as regime, a recent paper has proposed using beams from plasma wakefield accelerators compressed to an ultra-short bunch length (22.3 nm FWHM) to generate sub-100 as TW-scale pulses at 10 nm wavelength \cite{Emma2021} . The working principle of this plasma-driven scheme takes advantage of the compressed beam's high peak current and natural bunching factor at tens of nanometers to reach saturation in an ultra-short m-length undulator \cite{Emma2022a, Hessami2022, Robles2022}. 









Harmonic up-conversion for FEL lasing has been demonstrated in Ref. \cite{Yu2000} using an external laser seed. Additional harmonic seeding methods include variations of High-Gain Harmonic Generation (HGHG) (EEHG) \cite{allaria2012highly, allaria2013two}, and Echo-Enabled Harmonic Generation \cite{stupakov2009using, xiang2012evidence, Hemsing2016, rebernik2019coherent}. Experiments with EEHG have shown that pulses at 5.9 nm with a FWHM between 130 and 175 fs can be generated using this technique \cite{rebernik2019coherent}. In this paper, we show that combining a plasma-driven attosecond X-ray source with a harmonic generation scheme enables the production of  high-power (GW-scale), sub-100-as X-ray pulses at wavelengths down to the sub-nm level with a 3.3 GeV energy electron bunch in an ultra-compact (m-length) undulator. 

The method described in Ref. \cite{Emma2021} can be extended to shorter wavelengths using a variation of the HGHG scheme, where the initial seed is provided by the coherent radiation generated by the nm-scale electron bunch from the plasma accelerator. In the harmonic generation scheme presented, the first undulator is used to both generate radiation at the fundamental wavelength while simultaneously increasing the beam's bunching factor at higher harmonics. Following this initial section the harmonic bunching is exploited in one or more downstream undulators to generate short wavelength radiation with a pre-bunched beam. Additionally, since these beams are pre-bunched and the current spike in the beam is attosecond-scale, the produced X-ray beams are also attosecond scale. In particular, this work presents two simulation studies: a harmonic generation configuration where the fundamental is tuned to 10 nm and the fifth harmonic at 2 nm is utilized, and another configuration where the fundamental is tuned to 2 nm and the fifth harmonic at 0.40 nm is utilized. Since the fifth harmonic is at 2 nm, which lies beyond the carbon, nitrogen and oxygen K-edges, this scheme may provide utility for probing a multitude of chemical and biological systems.


\subsection{Theoretical background}


Typical FEL operation relies on the Self-Amplified Spontaneous Emission (SASE) process to generate radiation \cite{Bonifacio1984}. The minimum pulse length at saturation in the SASE process is set by the FEL cooperation length $L_c = \lambda/4\sqrt3\pi \rho$ where $\lambda$ is the desired photon wavelength and $\rho$ is the FEL parameter, typically $\mathcal{O}(10^{-4}-10^{-3})$ at X-ray wavelengths. This minimum pulse length, much longer than the photon wavelength, can be made shorter by increasing the magnitude of the FEL parameter using a high current spike to drive the FEL process. This was originally proposed in a scheme known as enhanced SASE in Ref. \cite{Zholents2005}. The original enhanced SASE scheme relies on using an external laser to chirp the beam before a weak bunch compression stage transforms the energy chirp into a large current spike at the compressor exit.
It was also shown that in an enhanced SASE experiment, longitudinal space-charge forces play a crucial role in re-shaping the phase-space of the lasing spike \cite{ding2009generation}, leading to the generation of pulses significantly shorter than the cooperation length \cite{baxevanis2018,Duris}
Recent experimental efforts have demonstrated this concept \cite{Duris2021} and variations of this idea \cite{Duris, Zhang2020} to successfully generate 100-as-scale duration XFEL pulses at nm wavelengths. 

\begin{figure} [!htb]
\begin{center}  
\includegraphics[width=\linewidth]{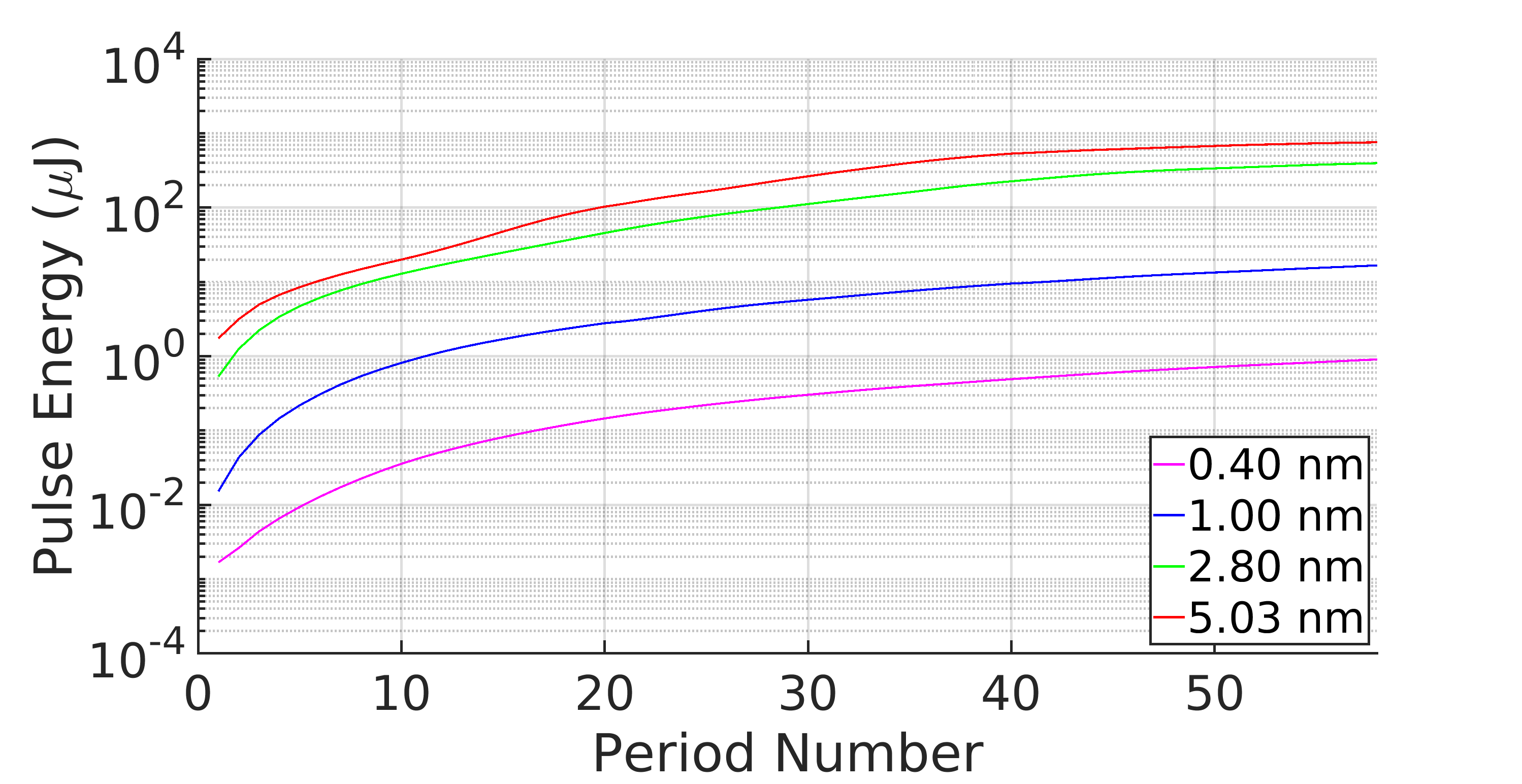}
\includegraphics[width=\linewidth]{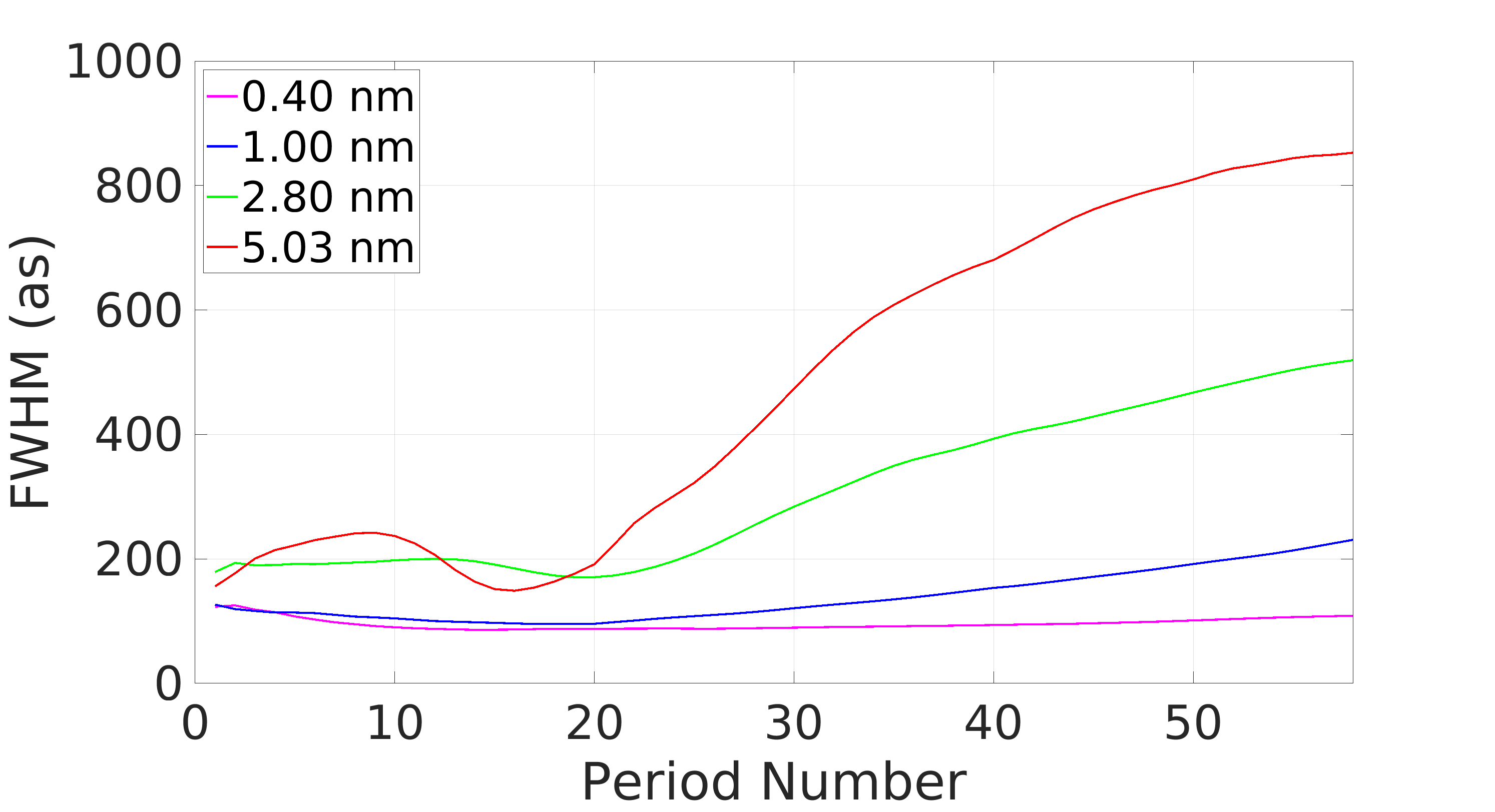}
\caption{X-ray pulse energy and pulse length along the undulator at different resonant wavelengths showing a decline in pulse energy and increase in saturation length as the resonant wavelength is scaled down towards 0.40 nm. The decline in pulse energy and generation of pulse lengths consistently over 100 as at shorter X-ray wavelength motivates the use of the harmonic generation scheme to scale a plasma-based attosecond X-ray source into nm wavelengths while maintaining sub-100-as pulse duration.}
\label{fig:power_comp}
\end{center}
\end{figure}

Scaling this approach to shorter pulse duration using a plasma accelerator was proposed in Ref. \cite{Emma2021}. The plasma-driven approach takes advantage of the high brightness beams with naturally strong energy chirps produced in plasma accelerators, which can be compressed in a weak downstream chicane to extreme, near-MA peak currents and ultra-short 10-nm-scale bunch lengths. While this approach is successful at generating high-power attosecond pulses at the 10 nm wavelength level, the pulse energy drops significantly when lasing directly at wavelengths in the nm-range, limited by the reduction in the beam's bunching factor at shorter wavelengths at the entrance to the undulator. Figure \ref{fig:power_comp} demonstrates how the X-ray pulse energy and pulse duration generated by directly sending the compressed beam from a plasma-based accelerator into an undulator decreases as the wavelength shortens (see table \ref{tab:onestage_sim_params} for simulation parameters). This is expected, due to the reduced  bunching factor and increased sensitivity to 3D effects such as emittance and energy spread at shorter wavelengths. In particular, Fig \ref{fig:power_comp} shows that the X-ray pulse energy drops by three orders of magnitude when scaling from 5.03 nm to 0.40 nm wavelength. Concurrently the minimum pulse length in this wavelength region is limited to 100 as or more.


\begin{table}[!htb]
\centering
\begin{tabular}{lll}
\hline
Resonant wavelength (nm)     & K Value & Undulator Period (cm)  \\ \hline
0.40                         & 0.315   & 3.00 \\
1.00                         & 0.980   & 5.60 \\
2.80                         & 2.510   & 5.60 \\
5.03                         & 3.600   & 5.60
\end{tabular}
\caption{Simulation parameters for a single undulator tuned to successively reach shorter resonant wavelengths in a plasma-driven attosecond XFEL.}
\label{tab:onestage_sim_params}
\end{table}

Motivated by these considerations, in this work, we investigate using a harmonic-based approach to generate sub-100-as X-ray pulses with a plasma-based attosecond XFEL. The beam is initially transported through a first undulator section tuned to a sub-harmonic of the desired output wavelength, generating radiation at the sub-harmonic while increasing the bunching at higher harmonics. Following the first section, the pre-bunched beam is transported into one or more downstream undulators tuned to the desired wavelengths. This results in the emission/generation of a two-color attosecond x-ray pulse pair with GW-power. In the following sections we discuss the two schemes using the same electron beam input parameters at the start of each of the two configurations as used in the single-stage simulations. 

\section{Harmonic generation simulations for a plasma-driven attosecond XFEL }

\begin{figure*}[!tbh]
    \centering
    \includegraphics*[width=\linewidth]{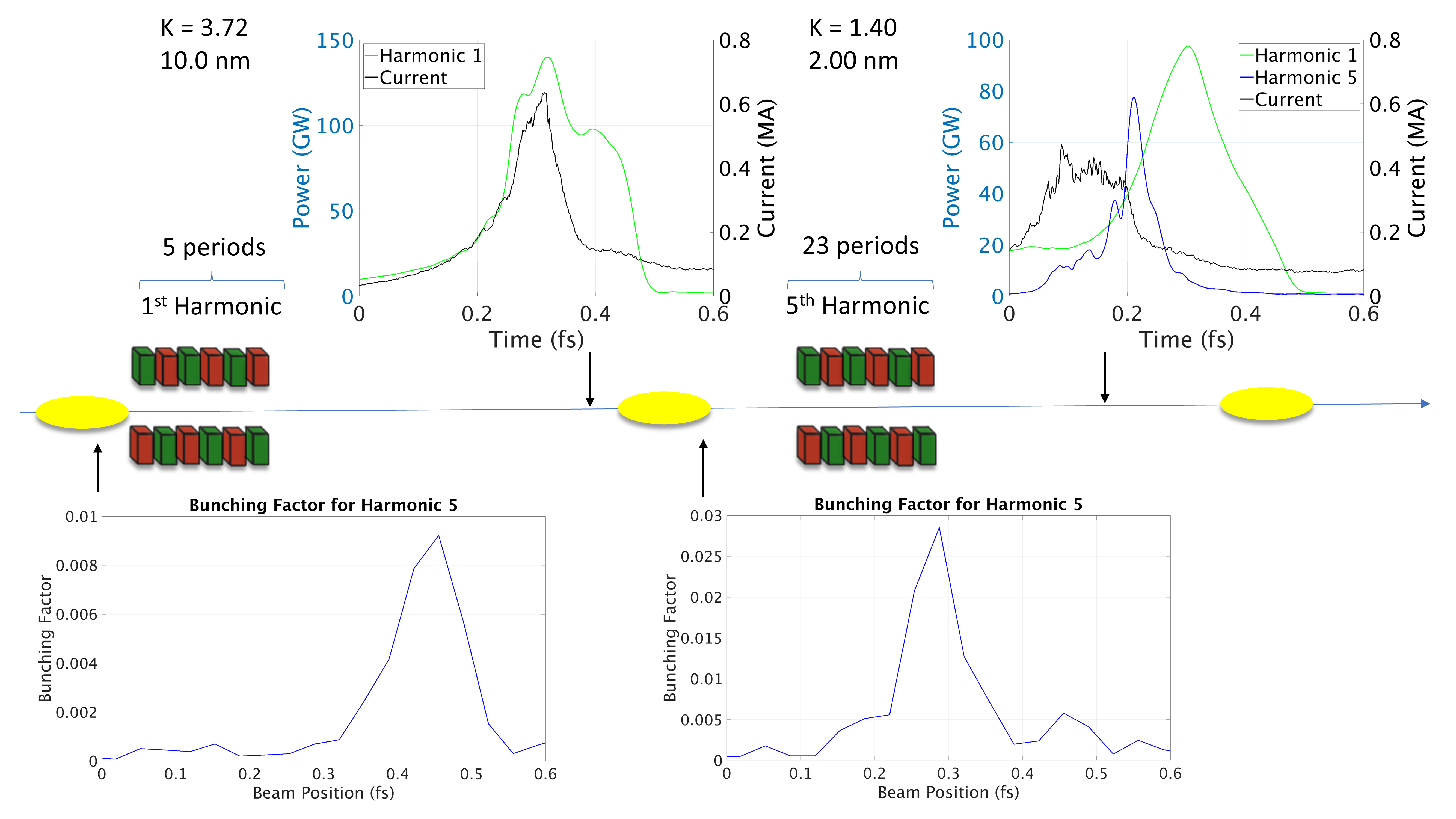}
    \caption{X-ray power profiles for the fundamental and fifth harmonic along the 2 nm attosecond beamline as well as the bunching factor for the fifth harmonic. This shows how a high-power, attosecond pulse at 2 nm can be produced compactly using Coherent Harmonic Generation and functions as an extension to the TW-scale 10 nm pulse proposed in Ref. \cite{Emma2021}. This photon energy extends beyond the C, N and O K-edges, making it useful for various chemical and biological systems \cite{Attwood1999}. The yellow circle represents the electron beam after it has been generated in a PWFA and fully compressed travelling down the set of undulators. }
    \label{fig:2nm_beamline}
\end{figure*}

The electron beam used in the harmonic generation simulations is the fully compressed attosecond duration bunch described in Ref. \cite{Emma2021} and obtained from Particle-In-Cell simulations of the plasma wakefield acceleration process along with beam dynamics simulations of the bunch compression stage following the plasma accelerator. The notable properties of this electron bunch are the extremely high peak current (0.714 MA) and ultra-short bunch duration at the current spike (22.3 nm FWHM) which result in a substantial (\%-level) fraction of pre-bunching at wavelengths down to 10 nm. Additionally, the beam has an energy of 3.28 GeV.

The FEL simulations using the electron beam as input are performed using the FEL code PUFFIN \cite{Campbell2012}. This FEL code is well suited for simulating multi-color FEL operation as well as FELs which use ultra-short beams on the order of an FEL cooperation length \cite{Jaroszynski1993} due to the fact that the code does not employ the Slowly Varying Envelope approximation (SVEA) nor the period averaging approximation. In addition, to simulate very high peak current beams such as the one considered in this work, the code has been modified to include the effects of longitudinal space charge (LSC) which introduce non-negligible dynamics and affect the beam distribution during its traversal in the undulator primarily through the generation of strong energy chirps in the longitudinal phase space. These chirps have known impacts on attosecond XFELs and are responsible for, among other factors, the shortening of the FEL pulse duration at saturation at the cost of reduced peak power \cite{baxevanis2018}. 


\subsection{2 nm Simulation}

The first harmonic generation scheme we consider is a two-stage undulator setup in which the first undulator is tuned to the fundamental at 10 nm, while the second undulator is tuned to 2 nm, the fifth harmonic. This builds on Ref. \cite{Emma2021}, where the wavelength is tuned to 10 nm. The parameters of the simulation are given in Table \ref{tab:2nm_sim_params}. The first undulator is very short, ending after 5 periods at 28 cm. This increases the bunching at the fifth harmonic from 0.9\% to a peak of 2.9\%, seeding the beam for its entry into the undulator tuned to the fifth harmonic, which consists of 23 periods with a total length of 1.29 m.  This distance was chosen empirically to maximize the bunching factor of the fifth harmonic while reducing the undulator length to deter excessive modulation of the phase space due to space charge effects and increasing the beam energy spread due to the FEL process. Since the source term for the FEL in pre-bunched FEL operation is proportional to the square of the bunching factor, this increase in bunching is equivalent to an increase in initial bunching equivalent seed power by almost an order of magnitude. The large bunching fraction at the entrance of the second undulator enables lasing at high power in a short undulator. The slippage between the radiation and the electron beam causes the attosecond pulse to lengthen. By radiating at the shorter wavelength, the slippage is reduced resulting in a shorter final pulse. 

\begin{table}[!htb]
\centering
\begin{tabular}{llll}
\hline
Pulse parameters               & Value               \\ \hline
Photon energy                & 0.620 keV              \\
Pulse length                 & 41.9 as              \\
Pulse energy                 & 6.85 $\mu$J                    \\
Peak power                   & 77.6 GW           \\ \hline
Simulation parameters                &               \\ \hline
Undulator period              & 5.60 cm           \\
Harmonic 1 K                  & 3.72             \\
Harmonic 1 Number of Periods  & 5                \\
Harmonic 1 Photon wavelength  & 10.00 nm              \\
Harmonic 5 K                  & 1.40             \\
Harmonic 5 Number of Periods  & 23                \\
Harmonic 5 Photon wavelength  & 2.00 nm              \\
\end{tabular}
\caption{Simulation parameters for the 2 nm harmonic generation setup.}
\label{tab:2nm_sim_params}
\end{table}

The beam dynamics in this multi-undulator harmonic generation setup are strongly affected by the LSC forces acting on the beam as it traverses the undulator. LSC can be neglected in the limit $L_p \ll L_g$, where $L_p$ is the relativistic plasma wavelength and $L_g$ is the FEL gain length. The relativistic plasma wavelength is given by equation \ref{eqn:plasmalambda}.

\begin{equation} \label{eqn:plasmalambda}
    L_p = \sqrt{\frac{I_A \gamma \gamma_z^2 \sigma_r^2}{2I_e}}
\end{equation}

$\sigma_r$ is the electron beam size, $I_A \approx 17 $ kA is the Alv\'{e}n current, and $I_e$ is the electron beam peak current \cite{Marcus2011}. 

Typical XFEL operational parameters satisfy this condition so LSC is usually a perturbation to the beam dynamics. However, the parameters used in this experiment produce $L_p$ and $L_g$ within a factor of 2 of each other, so this assumption is no longer valid. For this reason, we added in space charge effects using the treatment in Ref \cite{Geloni2007} to the PUFFIN code. We apply the energy modulation due to the longitudinal space charge at each undulator period and this contributes to the increase of non-negligible energy chirps on the electron beam as it traverses the undulator. In the case of the 2 nm simulation, the chirp increases in the first undulator at an average rate of 0.511 MeV/nm per period.

The main effect of LSC here is to introduce a positive energy chirp on the beam, lengthening the bunch as it traverses the undulator and reducing the peak current. This effect further underscores the importance of using short undulators in this harmonic-generating plasma-driven XFEL scheme to drive pre-bunching and X-ray pulse generation before spoiling the beam quality and degrading FEL performance in downstream sections.

\begin{figure} [t]
    \centering
    \includegraphics[width=\linewidth]{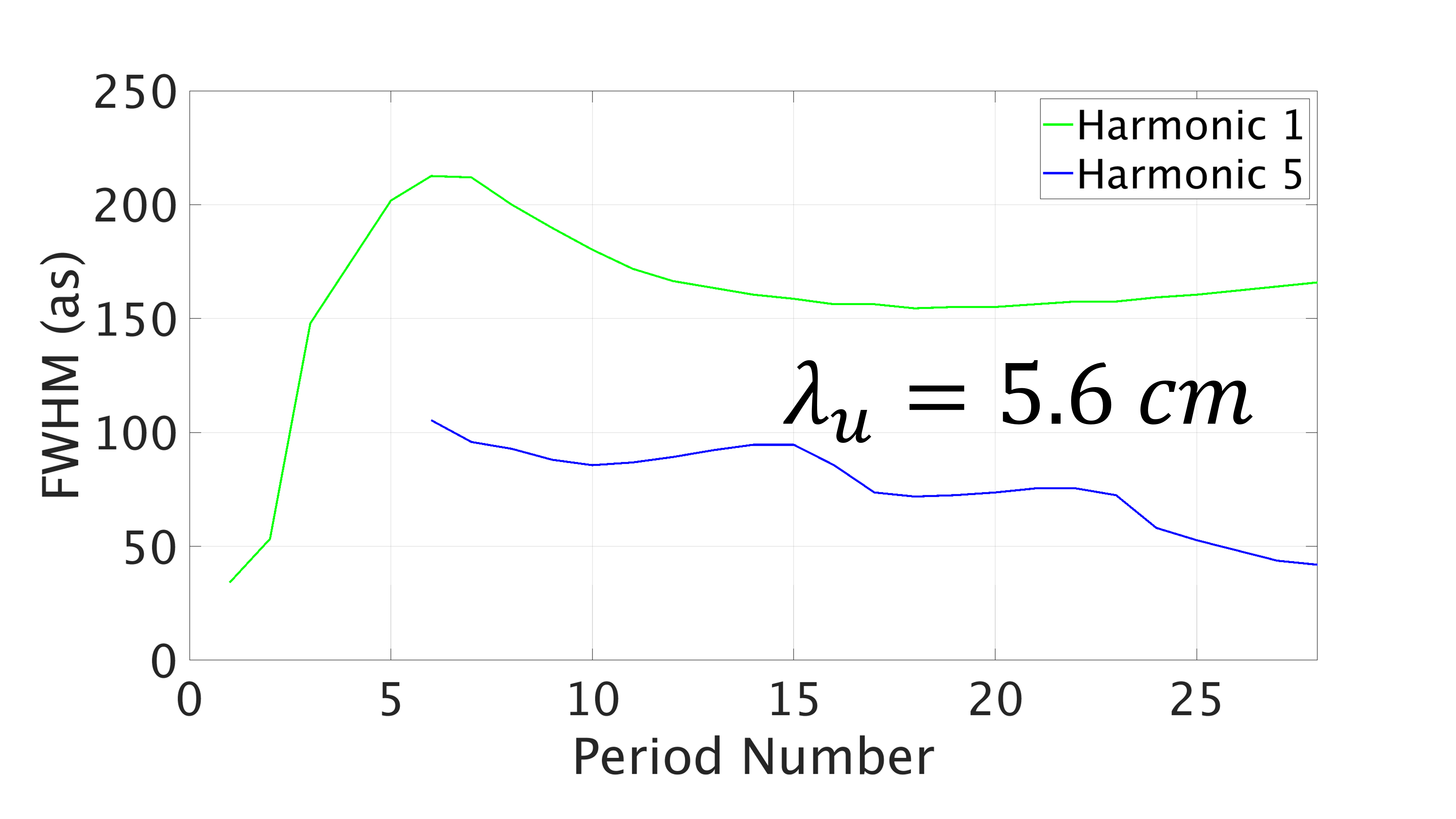}
    \caption{Pulse length produced in the 2 nm simulation. The minimum pulse length here is competitive with the shortest HHG pulses ever measured, 43 as \cite{Gaumnitz2017}.}
    \label{fig:2nm_FWHM}
\end{figure}

Figure \ref{fig:2nm_beamline} shows the longitudinal power profile of the two colors along the beamline in the 2 nm simulation. The spike in the bunching factor is responsible for the 2 nm pulse produced at the end of this beamline. Along the beamline, as the radiation slips out of the current spike, the region of bunching drifts away from the pulse, which would produce more radiation spikes separated in time. For that reason, the second undulator stage was chosen to cut off before this affect can arise, mitigating the generation of a multi-cycle pulse and increasing the pulse length. Figure \ref{fig:2nm_FWHM} depicts the FWHM of the pulses produced by the setup described in figure \ref{fig:2nm_beamline}. The final pulse length for the original 10 nm pulse is  166 as, while the 2 nm pulse leaves the beamline with a final pulse length of 41.9 as. Table \ref{tab:2nm_sim_params} summarizes the final photon pulses produced by this scheme.

We note that such short pulses are at the limit of what has been achieved to date in terms of minimum pulse duration with High-Harmonic Generation (HHG) sources \cite{Gaumnitz2017}. In contrast with HHG sources, the much higher ($\sim 10^6$x) pulse energy of this attosecond XFEL pulse enables the study of nonlinear processes with sub-100-as resolution as well as the application of two-color attosecond pump-probe techniques due to the multicolor nature of the output.

\subsection{0.40 nm Simulation}

\begin{figure*}[!tbh]
    \centering
    \includegraphics*[width=\linewidth]{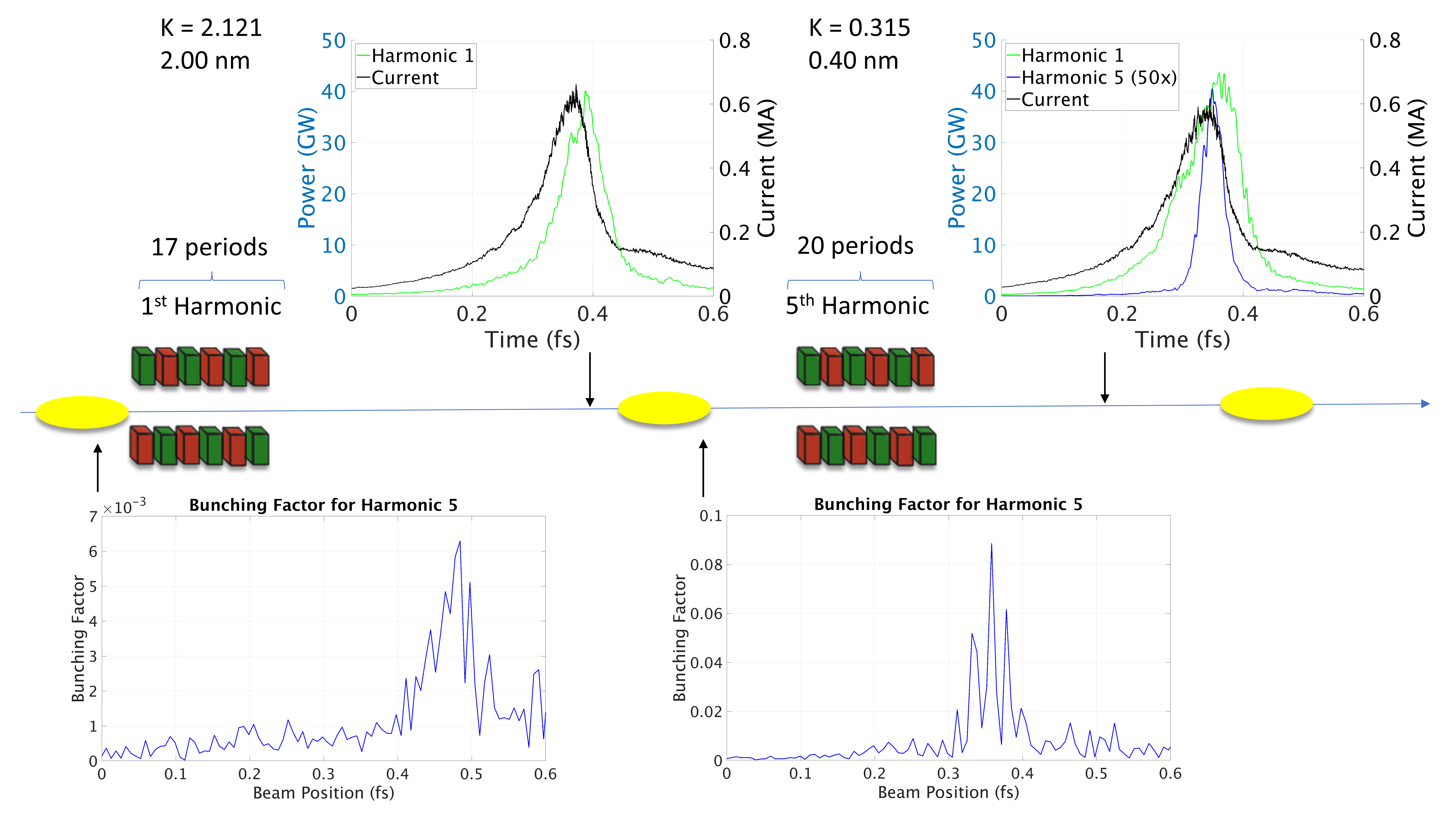}
    \caption{X-ray power profiles for both the fundamental and fifth harmonic along the two-stage attosecond harmonic-generating beamline tuned to 2 nm and 0.40 nm, as well as the bunching factor for the fifth harmonic showing an order of magnitude increase before and after the first undulator section.}
    \label{fig:0p4nm_beamline}
\end{figure*}

The second harmonic generation scheme we consider is a two-stage undulator setup  in which the first undulator is tuned to the fundamental at 2 nm, while the second undulator is tuned to 0.40 nm. These wavelengths were chosen to demonstrate how harmonic generation can be used to produce a 100-as or shorter X-ray pulse at a photon wavelength shorter than possible by conventional means. Other methods to produce GW or TW X-ray pulses using conventional FEL technologies produce pulses significantly longer, over 100-as, pulses \cite{huang2017generating, prat2023}. The parameters of the undulator are given in table \ref{tab:0p4nm_sim_params}. The first undulator ends after 17 periods at 51 cm, chosen much like the 2 nm simulation to maximize the growth of the fifth harmonic bunching factor and minimize excessive modulation due to space charge. The increase in chirp in the first undulator is significantly smaller than in the 2 nm simulation, at an average of 0.0921 MeV/nm per period, which is why the effect on the current spike is more pronounced in the 2 nm simulation, whereas the current profile maintains its structure over more periods in the 0.4 nm simulation. Figure \ref{fig:0p4nm_beamline} shows the longitudinal power profile of each color along the beamline in the two-stage simulation.

\begin{table}[b]
\centering
\begin{tabular}{llll}
\hline
Pulse parameters               & Value               \\ \hline
Photon energy                & 3.44 keV              \\
Pulse length                 & 40.2 as              \\
Pulse energy                 & 57.6 nJ                    \\
Peak power                   & 0.808 GW           \\ \hline
Simulation parameters                &               \\ \hline
Undulator period              & 0.03 m           \\
Harmonic 1 K                  & 2.121             \\
Harmonic 1 Number of Periods  & 17                \\
Harmonic 1 Photon wavelength  & 2 nm              \\
Harmonic 5 K                  & 0.315             \\
Harmonic 5 Number of Periods  & 20                \\
Harmonic 5 Photon wavelength  & 0.40 nm              \\
\end{tabular}
\caption{Simulation parameters for the 0.4 nm harmonic generation setup.}
\label{tab:0p4nm_sim_params}
\end{table}

\begin{figure} [!htb]
    \centering
    \includegraphics[width=\linewidth]{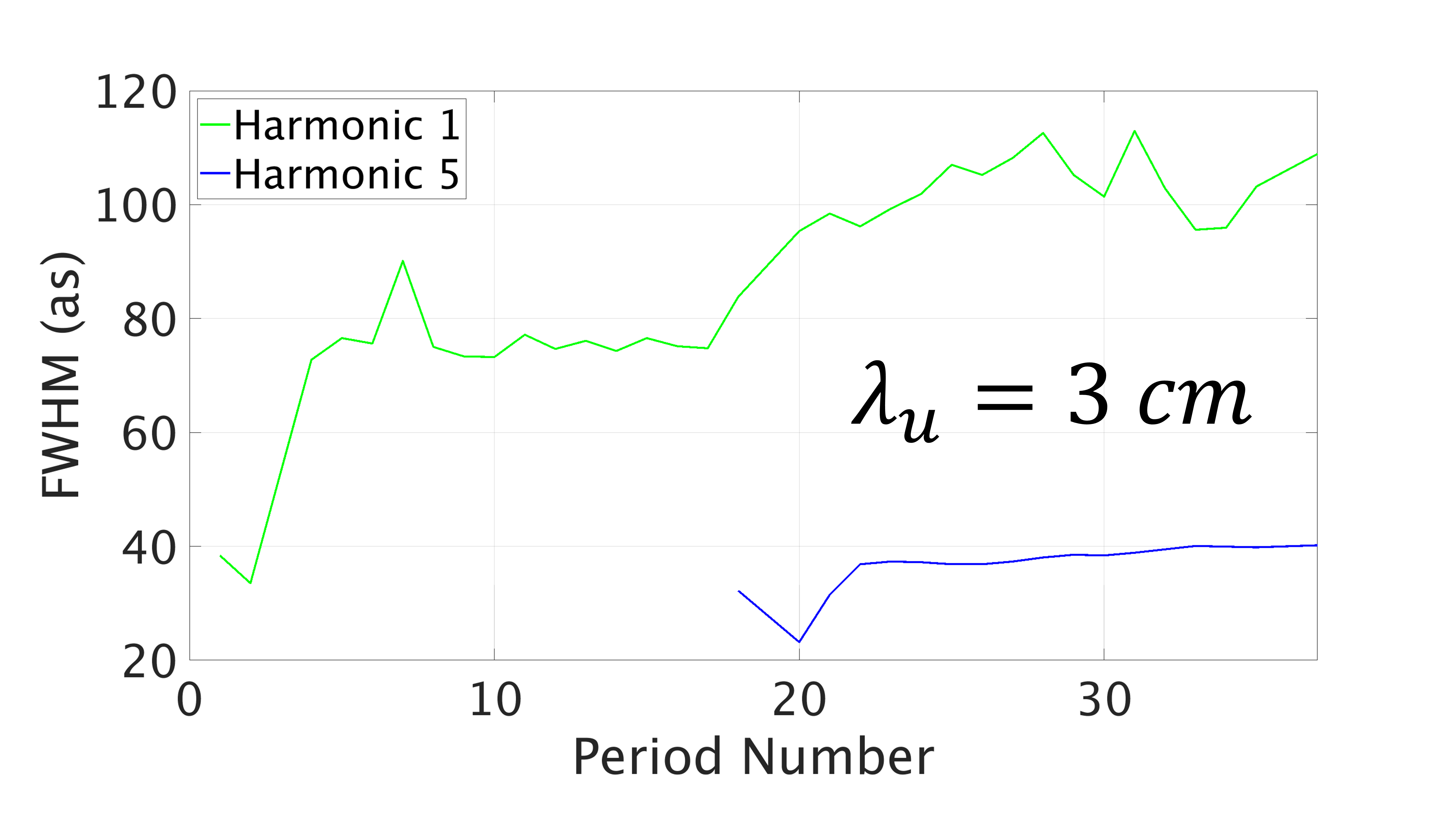}
    \caption{Pulse length produced in the 0.40 nm simulation. This demonstrates how the pulse length at the fifth harmonic remains sub-100-as across the entire beamline.}
    \label{fig:0p4nm_FWHM}
\end{figure}

The pulse envelope of the fifth harmonic is also delayed by 48 as behind the fundamental and closely resembles the profile of the current spike, further indicating this pulse was produced in the second undulator, utilizing the bunching produced in the first undulator. Additionally, it reaches a high peak power of 0.808 GW. The growth in the bunching factor in figure \ref{fig:0p4nm_beamline} also highlights the utility of this method. The bunching factor grows from a peak of  0.7\% to 9\%.  Figure \ref{fig:0p4nm_FWHM} depicts how this pulse length evolves over the course of the beamline, showing it stabilize at 40.2 as FWHM. Table \ref{tab:0p4nm_sim_params} summarizes the key parameters of the final photon pulse generated and the simulation parameters in the 0.4 nm case.

\section{Conclusion}

In this paper, we have presented a method for reaching short sub-nm wavelengths and multi-color operation in a plasma driven attosecond X-ray source. The scheme exploits harmonic generation in a multi-undulator setup to generate high harmonic bunching and employ retuned undulators to convert the bunching into harmonic radiation. This approach enables a new frontier of attosecond X-ray science through the production of ultra-short (sub-100-as) high peak power (GW-level) pulses with wavelength tunability and at wavelengths as short as 0.40 nm exceeding the Carbon K-edge, as well as the K-edges for most biologically relevant elements. The simulation studies presented have demonstrated the ability to use this technique in two largely interchangeable setups: one to produce two pulses at tens of gigawatts of power and wavelengths at 10 nm and 2 nm, useful for pump-probe experiments of materials with K-edges above 2 nm, and one to produce two as pulses at 2 nm and 0.4 nm, useful for pump-probe applications where keV-scale photon energies, GW-scale powers, and attosecond pulses are necessary. We note that different undulator tapering strategies can be utilized in conjunction with the harmonic-based approach presented here to, for example, increase the radiation bandwidth, maximize bunching while reducing fundamental output, or maximize photon production in a given section to tailor the pulse properties for particular experiments. These manipulations will be investigated in future work.


\begin{acknowledgments}
This work was supported by the U.S. Department of Energy under contract number DE-AC02-76SF00515.
\end{acknowledgments}

\appendix


\bibliography{CHGpaper}
\bibliographystyle{ieeetr}

\end{document}